\documentclass[sn-mathphys]{sn-jnl}

\jyear{2021}%

\theoremstyle{thmstyleone}%
\theoremstyle{thmstyletwo}%

\theoremstyle{thmstylethree}%

\usepackage{comment}
\usepackage{color}
\usepackage{caption}
\begin{document}

\title[IXPE Observations of the Crab Pulsar and Nebula]{Simultaneous space and phase resolved X-ray polarimetry of the Crab Pulsar and Nebula}


\author*[1,2,3]{\fnm{Niccol\`o} \sur{Bucciantini}}\email{niccolo.bucciantini@inaf.it}

\author[1]{\fnm{Riccardo} \sur{Ferrazzoli}}
\author[5]{\fnm{Matteo} \sur{Bachetti}}
\author[4]{\fnm{John} \sur{Rankin}}
\author[6]{\fnm{Niccol\`o} \sur{Di Lalla}}
\author[7]{\fnm{Carmelo} \sur{Sgr\`o}}
\author[6]{\fnm{Nicola} \sur{Omodei}}
\author[8]{\fnm{Takao} \sur{Kitaguchi}}
\author[9]{\fnm{Tsunefumi} \sur{Mizuno}}
\author[10]{\fnm{Shuichi} \sur{Gunji}}
\author[10]{\fnm{Eri} \sur{Watanabe}}
\author[7,11]{\fnm{Luca} \sur{Baldini}}
\author[12]{\fnm{Patrick} \sur{Slane}}
\author[13]{\fnm{Martin~C.} \sur{Weisskopf}}
\author[6]{\fnm{Roger~W.} \sur{Romani}}
\author[5]{\fnm{Andrea} \sur{Possenti}}
\author[14]{\fnm{Herman~L.} \sur{Marshall}}
\author[7]{\fnm{Stefano} \sur{Silvestri}}
\author[4]{\fnm{Luigi} \sur{Pacciani}}
\author[15,16,17]{\fnm{Michela} \sur{Negro}}
\author[4]{\fnm{Fabio} \sur{Muleri}}
\author[18]{\fnm{Emma} \sur{de O\~na Wilhelmi}}
\author[19,4]{\fnm{Fei} \sur{Xie}}
\author[20]{\fnm{Jeremy} \sur{Heyl}}
\author[7]{\fnm{Melissa} \sur{Pesce-Rollins}}
\author[6]{\fnm{Josephine} \sur{Wong}}
\author[5]{\fnm{Maura} \sur{Pilia}}
\author[21]{\fnm{Iv\'an} \sur{Agudo}}
\author[22]{\fnm{Lucio~A.} \sur{Antonelli}}
\author[13]{\fnm{Wayne~H.} \sur{Baumgartner}}
\author[7]{\fnm{Ronaldo} \sur{Bellazzini}}
\author[24]{\fnm{Stefano} \sur{Bianchi}}
\author[13]{\fnm{Stephen~D.} \sur{Bongiorno}}
\author[25,26]{\fnm{Raffaella} \sur{Bonino}}
\author[7]{\fnm{Alessandro} \sur{Brez}}
\author[4]{\fnm{Fiamma} \sur{Capitanio}}
\author[7]{\fnm{Simone} \sur{Castellano}}
\author[27]{\fnm{Elisabetta} \sur{Cavazzuti}}
\author[28,23]{\fnm{Stefano} \sur{Ciprini}}
\author[4]{\fnm{Enrico} \sur{Costa}}
\author[4]{\fnm{Alessandra} \sur{De Rosa}}
\author[4]{\fnm{Ettore} \sur{Del Monte}}
\author[27]{\fnm{Laura} \sur{Di Gesu}}
\author[4]{\fnm{Alessandro} \sur{Di Marco}}
\author[27]{\fnm{Immacolata} \sur{Donnarumma}}
\author[29]{\fnm{Victor} \sur{Doroshenko}}
\author[31]{\fnm{Michal} \sur{Dov\v{c}iak}}
\author[13]{\fnm{Steven~R.} \sur{Ehlert}}
\author[8]{\fnm{Teruaki} \sur{Enoto}}
\author[4]{\fnm{Yuri} \sur{Evangelista}}
\author[4]{\fnm{Sergio} \sur{Fabiani}}
\author[32]{\fnm{Javier~A.} \sur{Garcia}}
\author[33]{\fnm{Kiyoshi} \sur{Hayashida}}
\author[34]{\fnm{Wataru} \sur{Iwakiri}}
\author[35,36]{\fnm{Svetlana~G.} \sur{Jorstad}}
\author[31]{\fnm{Vladimir} \sur{Karas}}
\author[13]{\fnm{Jeffery~J.} \sur{Kolodziejczak}}
\author[37]{\fnm{Henric} \sur{Krawczynski}}
\author[4]{\fnm{Fabio} \sur{La Monaca}}
\author[25]{\fnm{Luca} \sur{Latronico}}
\author[38]{\fnm{Ioannis} \sur{Liodakis}}
\author[25]{\fnm{Simone} \sur{Maldera}}
\author[7]{\fnm{Alberto} \sur{Manfreda}}
\author[39]{\fnm{Fr\'ed\'eric} \sur{Marin}}
\author[27]{\fnm{Andrea} \sur{Marinucci}}
\author[35]{\fnm{Alan~P.} \sur{Marscher}}
\author[25,26]{\fnm{Francesco} \sur{Massaro}}
\author[24]{\fnm{Giorgio} \sur{Matt}}
\author[40]{\fnm{Ikuyuki} \sur{Mitsuishi}}
\author[41]{\fnm{C.-Y.} \sur{Ng}}
\author[13]{\fnm{Stephen~L.} \sur{O’Dell}}
\author[25]{\fnm{Chiara} \sur{Oppedisano}}
\author[22]{\fnm{Alessandro} \sur{Papitto}}
\author[42]{\fnm{George~G.} \sur{Pavlov}}
\author[6]{\fnm{Abel~L.} \sur{Peirson}}
\author[22,23]{\fnm{Matteo} \sur{Perri}}
\author[43]{\fnm{Pierre-Olivier} \sur{Petrucci}}
\author[44,30]{\fnm{Juri} \sur{Poutanen}}
\author[23]{\fnm{Simonetta} \sur{Puccetti}}
\author[13]{\fnm{Brian~D.} \sur{Ramsey}}
\author[4]{\fnm{Ajay} \sur{Ratheesh}}
\author[4]{\fnm{Paolo} \sur{Soffitta}}
\author[7]{\fnm{Gloria} \sur{Spandre}}
\author[8]{\fnm{Toru} \sur{Tamagawa}}
\author[45]{\fnm{Fabrizio} \sur{Tavecchio}}
\author[46]{\fnm{Roberto} \sur{Taverna}}
\author[40]{\fnm{Yuzuru} \sur{Tawara}}
\author[13]{\fnm{Allyn~F.} \sur{Tennant}}
\author[13]{\fnm{Nicolas~E.} \sur{Thomas}}
\author[47,28,17]{\fnm{Francesco} \sur{Tombesi}}
\author[5]{\fnm{Alessio} \sur{Trois}}
\author[44,30]{\fnm{Sergey} \sur{Tsygankov}}
\author[46,48]{\fnm{Roberto} \sur{Turolla}}
\author[49]{\fnm{Jacco} \sur{Vink}}
\author[48]{\fnm{Kinwah} \sur{Wu}}
\author[48]{\fnm{Silvia} \sur{Zane}}


\affil[1]{INAF Osservatorio Astrofisico di Arcetri,\orgaddress{ Largo Enrico Fermi 5, 50125 Firenze, Italy}}
\affil[2]{Dipartimento di Fisica e Astronomia, Università degli Studi di Firenze,\orgaddress{ Via Sansone 1, 50019 Sesto Fiorentino (FI), Italy}}
\affil[3]{Istituto Nazionale di Fisica Nucleare, Sezione di Firenze, \orgaddress{Via Sansone 1, 50019 Sesto Fiorentino (FI), Italy}}
\affil[4]{INAF Istituto di Astrofisica e Planetologia Spaziali,\orgaddress{ Via del Fosso del Cavaliere 100, 00133 Roma, Italy}}
\affil[5]{INAF Osservatorio Astronomico di Cagliari,\orgaddress{Via della Scienza 5, 09047 Selargius (CA), Italy}}
\affil[6]{Department of Physics and Kavli Institute for Particle Astrophysics and Cosmology, Stanford University,\orgaddress{ Stanford, California 94305, USA.}}
\affil[7]{Istituto Nazionale di Fisica Nucleare, Sezione di Pisa,\orgaddress{Largo B. Pontecorvo 3, 56127 Pisa, Italy}}
\affil[8]{RIKEN Cluster for Pioneering Research,\orgaddress{ 2-1 Hirosawa, Wako, Saitama 351-0198, Japan }}
\affil[9]{Hiroshima Astrophysical Science Center, Hiroshima University,\orgaddress{1-3-1 Kagamiyama, Higashi-Hiroshima, Hiroshima 739-8526, Japan }}
\affil[10]{Yamagata University,\orgaddress{ 1-4-12 Kojirakawa-machi, Yamagata-shi 990-8560, Japan}}
\affil[11]{Dipartimento di Fisica, Università di Pisa; Largo B. Pontecorvo 3, 56127 Pisa, Italy}
\affil[12]{Center for Astrophysics, Harvard Smithsonian; 60 Garden St, Cambridge, MA 02138, USA}
\affil[13]{NASA Marshall Space Flight Center; Huntsville, AL 35812, USA}
\affil[14]{MIT Kavli Institute for Astrophysics and Space Research, Massachusetts Institute of Technology; 77 Massachusetts Avenue, Cambridge, MA 02139, USA}
\affil[15]{Center for Research and Exploration in Space Science and Technology, NASA/GSFC; Greenbelt, MD 20771, USA}
\affil[16]{NASA Goddard Space Flight Center; Greenbelt, MD 20771, USA}
\affil[17]{Department of Astronomy, University of Maryland; College Park, Maryland 20742, USA.}
\affil[18]{Deutsches Elektronen-Synchrotron (DESY), 15738 Zeuthen, Germany}
\affil[19]{Guangxi Key Laboratory for Relativistic Astrophysics, School of Physical Science and Technology; Guangxi University, Nanning 530004, China}
\affil[20]{University of British Columbia; Vancouver, BC V6T 1Z4, Canada}
\affil[21]{Instituto de Astrof\'isicade Andaluc\'ia, IAA-CSIC; Glorieta de la Astronom\'ia s/n, 18008 Granada, Spain}
\affil[22]{INAF Osservatorio Astronomico di Roma, \orgaddress{Via Frascati 33, 00078 Monte Porzio Catone, Roma, Italy}}
\affil[23]{Space Science Data Center, Agenzia Spaziale Italiana,\orgaddress{Via del Politecnico snc, 00133 Roma, Italy}}
\affil[24]{Dipartimento di Matematica e Fisica, Università degli Studi Roma Tre; Via della Vasca Navale 84, 00146 Roma, Italy}
\affil[25]{Istituto Nazionale di Fisica Nucleare, Sezione di Torino; Via Pietro Giuria 1, 10125 Torino, Italy}
\affil[26]{Dipartimento di Fisica, Universit\`a degli Studi di Torino; Via Pietro Giuria 1, 10125 Torino, Italy}
\affil[27]{ASI - Agenzia Spaziale Italiana; Via del Politecnico snc 00133 Roma, Italy}
\affil[28]{Istituto Nazionale di Fisica Nucleare, Sezione di Roma “Tor Vergata”; Via della Ricerca Scientifica 1, 00133 Roma, Italy}
\affil[29]{Institut für Astronomie und Astrophysik, Universität Tübingen; Sand 1, 72076 Tübingen, Germany}
\affil[30]{Space Research Institute of the Russian Academy of Sciences; Profsoyuznaya Str. 84/32, Moscow 117997, Russia}
\affil[31]{Astronomical Institute of the Czech Academy of Sciences; Bočnı II 1401/1, 14100 Praha 4, Czech Republic}
\affil[32]{California Institute of Technology; Pasadena, CA 91125, USA}
\affil[33]{Osaka University; 1-1 Yamadaoka, Suita, Osaka 565-0871, Japan}
\affil[34]{Department of Physics, Faculty of Science and Engineering, Chuo University; 1-13-27 Kasuga, Bunkyo-ku, Tokyo 112-8551, Japan}
\affil[35]{Institute for Astrophysical Research, Boston University; 725 Commonwealth Avenue, Boston, MA 02215, USA}
\affil[36]{Department of Astrophysics, St. Petersburg State University; Universitetsky pr. 28, Petrodvoretz, 198504 St. Petersburg, Russia}
\affil[37]{Physics Department and McDonnell Center for the Space Sciences; Washington University in St. Louis, St. Louis, MO 63130, USA}
\affil[38]{Finnish Centre for Astronomy with ESO; 20014 University of Turku, Finland}
\affil[39]{Université de Strasbourg, CNRS, Observatoire Astronomique de Strasbourg; UMR 7550, 67000 Strasbourg, France}
\affil[40]{Graduate School of Science, Division of Particle and Astrophysical Science; Nagoya University, Furo-cho, Chikusa-ku, Nagoya, Aichi 464-8602, Japan}
\affil[41]{Department of Physics, The University of Hong Kong; Pokfulam, Hong Kong}
\affil[42]{Department of Astronomy and Astrophysics, Pennsylvania State University; University Park, PA 16802, USA}
\affil[43]{Université Grenoble Alpes, CNRS, IPAG; 38000 Grenoble, France}
\affil[44]{Department of Physics and Astronomy, 20014 University of Turku; Finland}
\affil[45]{INAF Osservatorio Astronomico di Brera; Via E. Bianchi 46, 23807 Merate (LC), Italy}
\affil[46]{Dipartimento di Fisica e Astronomia, Università degli Studi di Padova; Via Marzolo 8,35131 Padova, Italy}
\affil[47]{Dipartimento di Fisica, Università degli Studi di Roma “Tor Vergata”; Via della Ricerca Scientifica 1, 00133 Roma, Italy}
\affil[48]{Mullard Space Science Laboratory, University College London; Holmbury St Mary, Dorking, Surrey RH5 6NT, UK}
\affil[49]{Anton Pannekoek Institute for Astronomy \& GRAPPA, University of Amsterdam; Science Park 904, 1098 XH Amsterdam, The Netherlands}


\abstract{The Crab pulsar and its nebula are among the most studied astrophysical systems, and constitute one of the most promising environments where high energy processes and particle acceleration can be investigated. They are the only objects for which previous X-ray polarisation has been reported. We present here the first Imaging X-ray Polarimetry Explorer (IXPE) observation of the Crab pulsar and nebula. The total pulsar pulsed emission in the [2–8]~keV energy range is unpolarised. Significant polarisation up to 15\% is detected only in the core of the main peak. The nebula has a total space integrated polarised degree of 20\% and polarisation angle of 145$^\circ$. The polarised maps show a large variation in the local polarisation, and regions with polarised degree up to 45-50\%. The polarisation pattern suggests a predominantly toroidal magnetic field.}

\keywords{X-rays: general, polarization, supernova remnants, pulsars: individual (Crab)}



\maketitle
\section{Introduction}\label{sec1}

The Crab pulsar (PSR B0531+21, PSR J0534+2200) and nebula (G184.6-5.8), born out of the supernova SN1054, form one of the most interesting high energy astrophysical systems, and one of the foremost environments where the physics of compact objects, particle acceleration, and relativistic outflows can be investigated \cite{Rees_Gunn74a,Gaensler_Slane06a,Hester08a,Buhler_Blandford14a}. \\
\\
The Crab pulsar (PSR) and nebula (PWN) are the only astrophysical system for which integrated and/or phase-resolved X-ray polarisation has been reported by various instruments \cite{Weisskopf_Silver+78a,Forot_Laurent+08a,Moran_Shearer+13b,Chauvin_Roques+13a,Chauvin_Floren+17a,Vadawale_Chattopadhyay+18a,Feng_Li+20a,Long_Feng+21a,Li_Produit+22a}, suggesting a polarised degree (PD) in the pulses $\sim$20\%, lower than in the interpulse. The integrated X-ray PD in the off pulse (OP) due to the PWN is consistently found to be $\sim$20\% with typical polarisation angle (PA) $\sim$130$^\circ$-140$^\circ$. The PWN polarimetry by OSO-8 has PD = 19.19\% $\pm$ 0.97\% and 19.50\% $\pm$ 2.77\% at 2.6~keV and 5.5~keV respectively, while PA = 156.36$^\circ$ $\pm$ 1.44$^\circ$ and 152.99$^\circ$ $\pm$ 4.04$^\circ$, at the same energies \cite{Weisskopf_Silver+78a} (all errors hereafter are at 1$\sigma$ confidence level, see also the Methods). \\
\\
The Crab PSR has a rotation period P = 33.7~ms, and spins down at a rate  $\dot{P} = 4.21\times10^{-13}$, corresponding to an effective surface dipole magnetic field of $3.8 \times 10^{12}$~G, and a spin-down luminosity  $\dot{E}$ = $4.3\times 10^{38 }$~erg~s$^{-1}$ \cite{Hester08a,Buhler_Blandford14a}. The estimated distance is 2~kpc \cite{Trimble73a}. Pulsed emission has been observed at all wavelengths from radio up to TeV. The pulse shape is characterized by a main pulse (P1), and an interpulse or second pulse (P2), whose positions in phase show only a slight variation with energy. At optical frequencies and above, a bridge (B) of emission is observed between them. The total unabsorbed pulsed X-ray luminosity in the [2-10]~keV band is $\sim 2.7 \times  10^{-9}$ erg~cm$^{-2}$~s$^{-1}$ \cite{Kaaret_Marshall+01a}, while the photon index in the energy band [1-100]~keV is found to be in the range 1.4-2.2 \cite{Massaro_Campana+06a,Weisskopf_Tennant+11a,Ge_Lu+12a,Vivekanand21a}. The photon index is phase dependent: the emission is harder for B than for the peaks, and slightly harder for P2 than P1. Optical and radio polarisation have been measured since the PSR discovery. The OP emission (in optical in the phase range 0.78-0.84) has a PD = 33\%, and PA = 130$^\circ$ \cite{Sowikowska_Kanbach+09a}. After OP subtraction the average PD of the pulses is found to be 5.5\% and the average PA is 96$^\circ$ \cite{Sowikowska_Kanbach+09a}.\\
\\
The Crab PWN has an apparent size of $6' \times 4'$ in the optical band, corresponding to about $11 \times 7$~ly at its estimated distance. It has a broad band non-thermal spectrum, extending from radio to TeV energies, due to synchrotron from high energy electrons and positrons up to $\sim$150~MeV, and Inverse Compton above. The [2-8]~keV integrated luminosity is $\sim1.8 \times 10^{-8}$~erg~cm$^{-2}$~s$^{-1}$ \cite{Kargaltsev_Pavlov08a} with a photon index $\sim2.12-2.15$ \cite{Mori_Burrows+04a}. X-ray imaging shows a well developed axisymmetric structure known as jet-torus \cite{Hester_Scowen+95a,Weisskopf_Hester+00a}. Radio polarisation shows typical average values of the PD in the range 5-10\%, and PA = 150$^\circ$, with no correlation with the X-ray features \cite{Aumont_Conversi+10a}. High resolution optical polarimetry of the inner region shows a PD = 40\% and PA = 127$^\circ$ \cite{Moran_Shearer+13a}.

\section{Results}\label{sec2}

\begin{figure}[h]%
\centering
\includegraphics[width=0.9\textwidth]{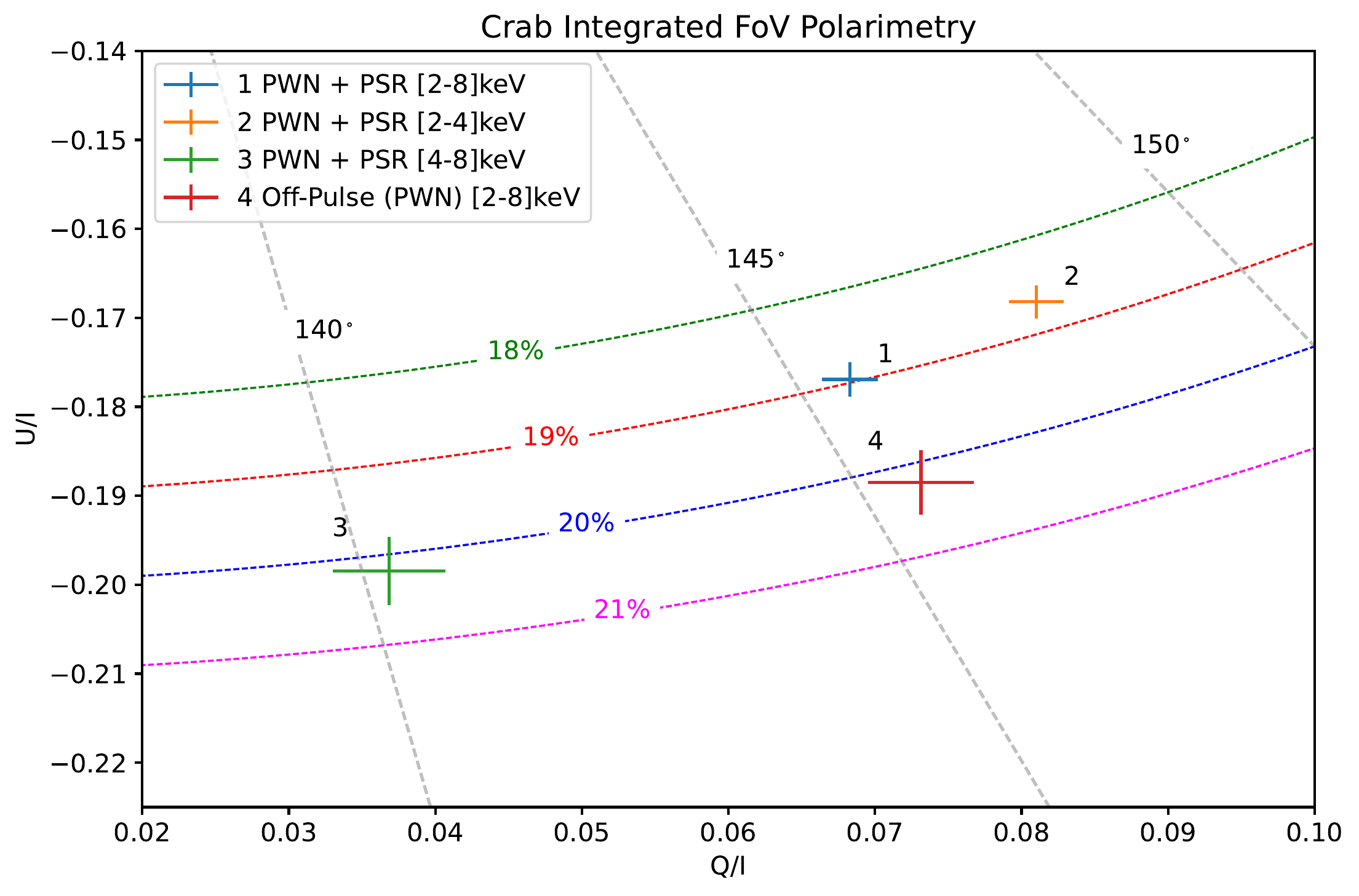}
\caption{\textbf{Global Polarisation properties of the Crab PWN$+$PSR complex.} Emission is integrated over a region of 2.5' radius centered on the PSR. Normalized Stokes parameters are shown for the total emission, the emission in the [2-4]~keV and [4-8]~keV energy bands, and for the OP only, together with contours of polarised degree (in \%) and angle.}\label{fig:1}
\end{figure}

\begin{figure}[h]%
\includegraphics[width=0.45\textwidth, trim = 0bp 0 0 0bp, clip]{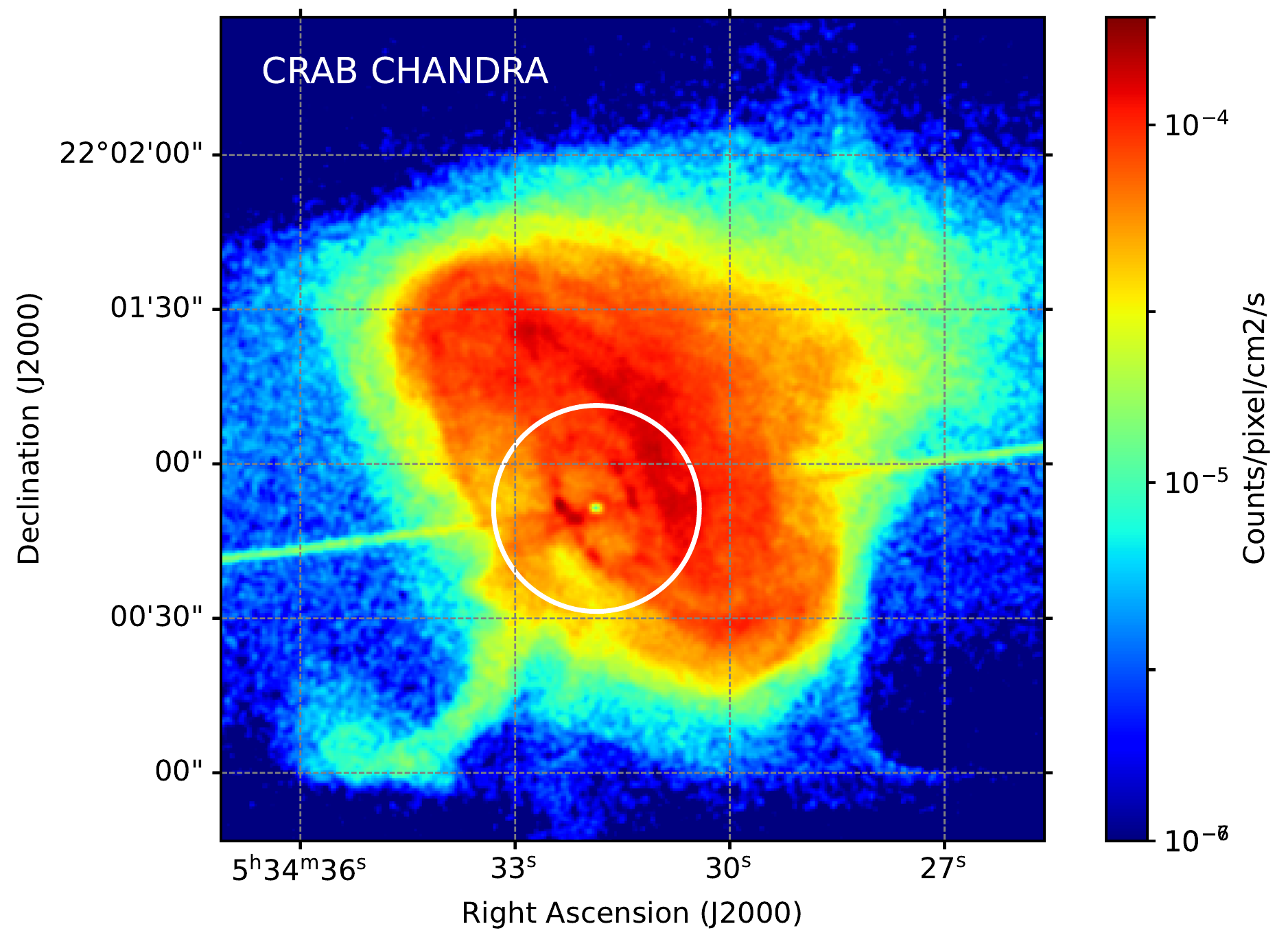}\hspace{0.6cm}
\includegraphics[width=0.40\textwidth, trim = 500 0 200 0, clip]{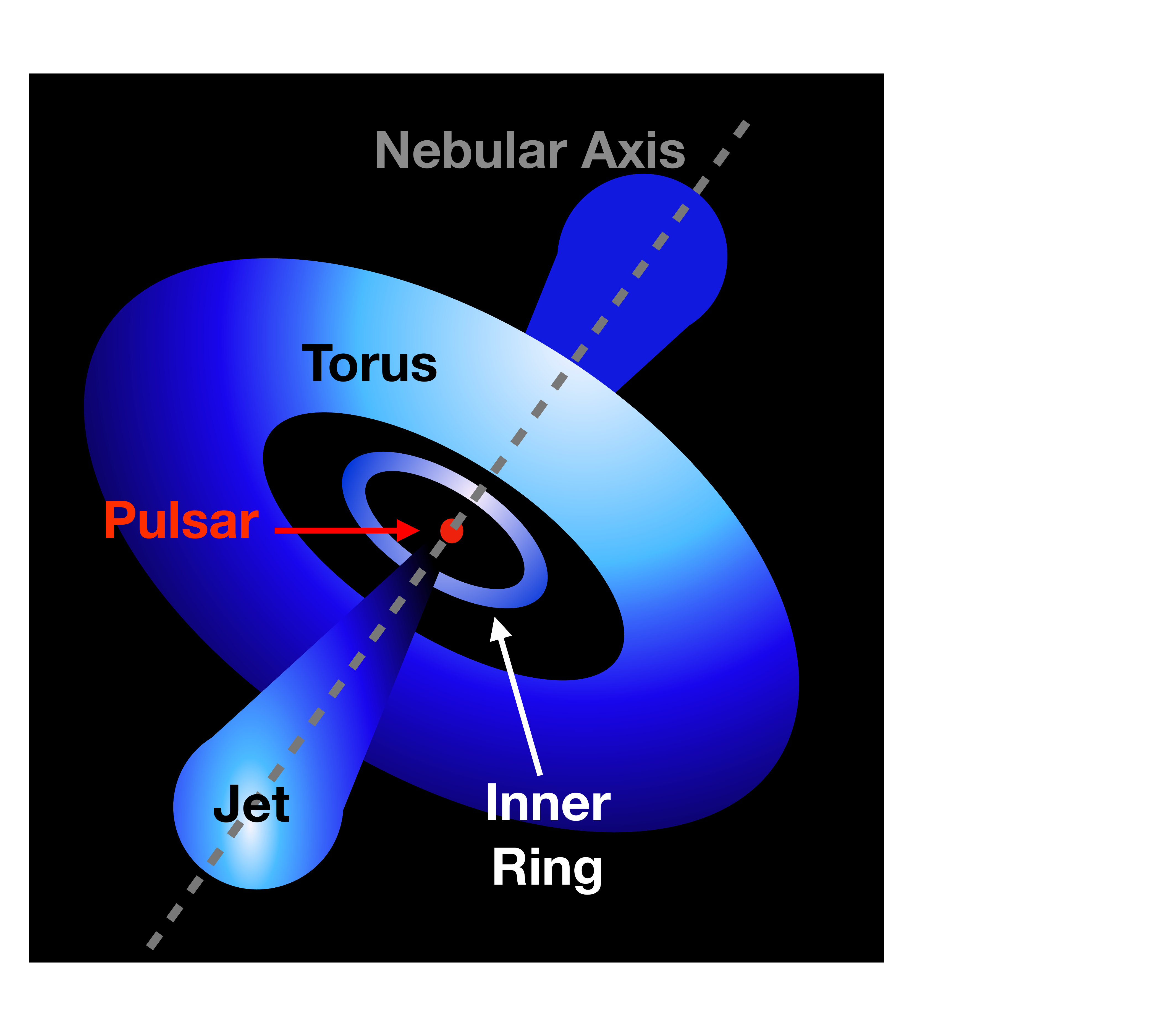}\hspace{2.cm}\\
\includegraphics[width=0.45\textwidth, trim = 0bp 0 0 0bp, clip]{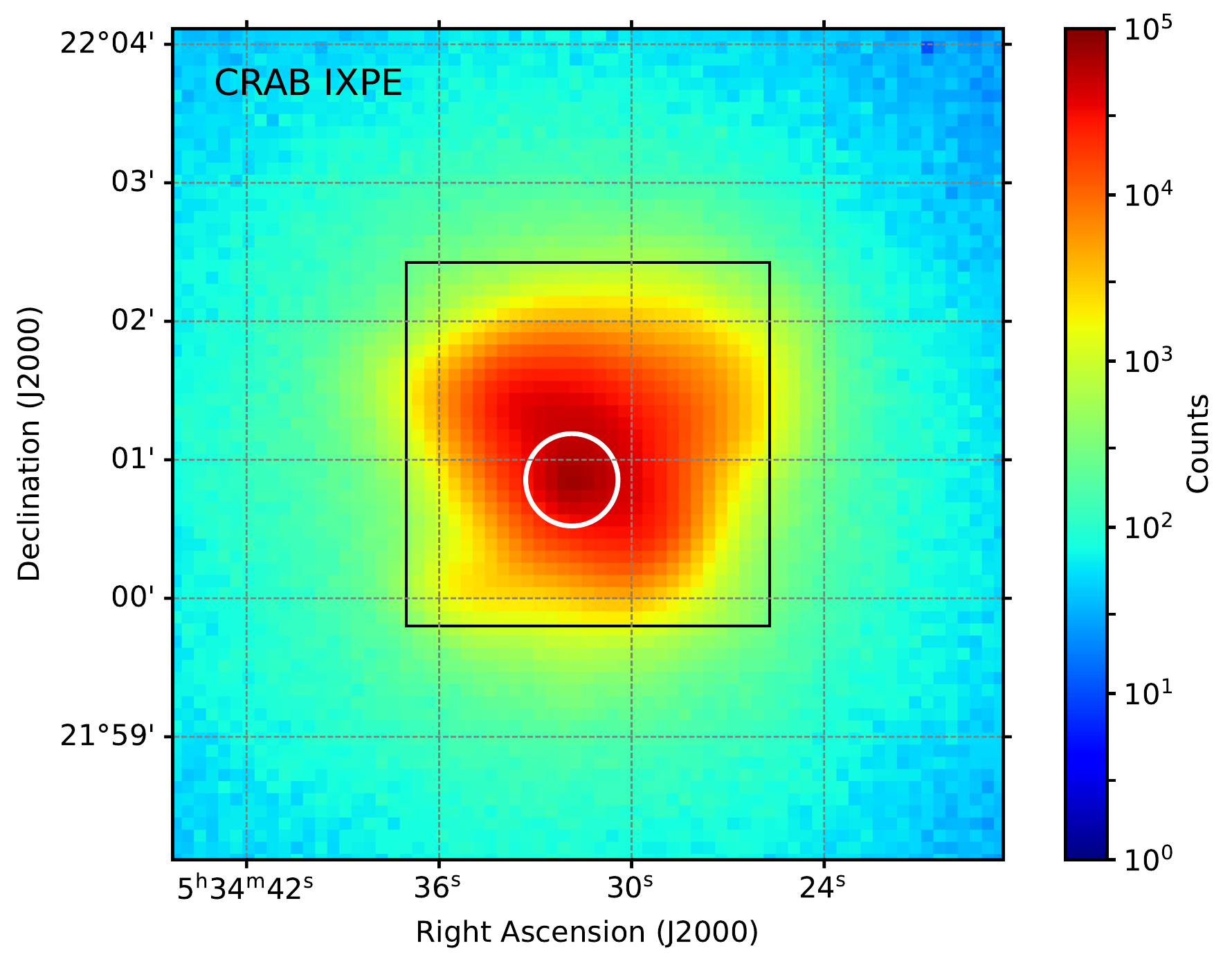}
\includegraphics[width=0.45\textwidth, trim = 0bp 0 0 0bp, clip]{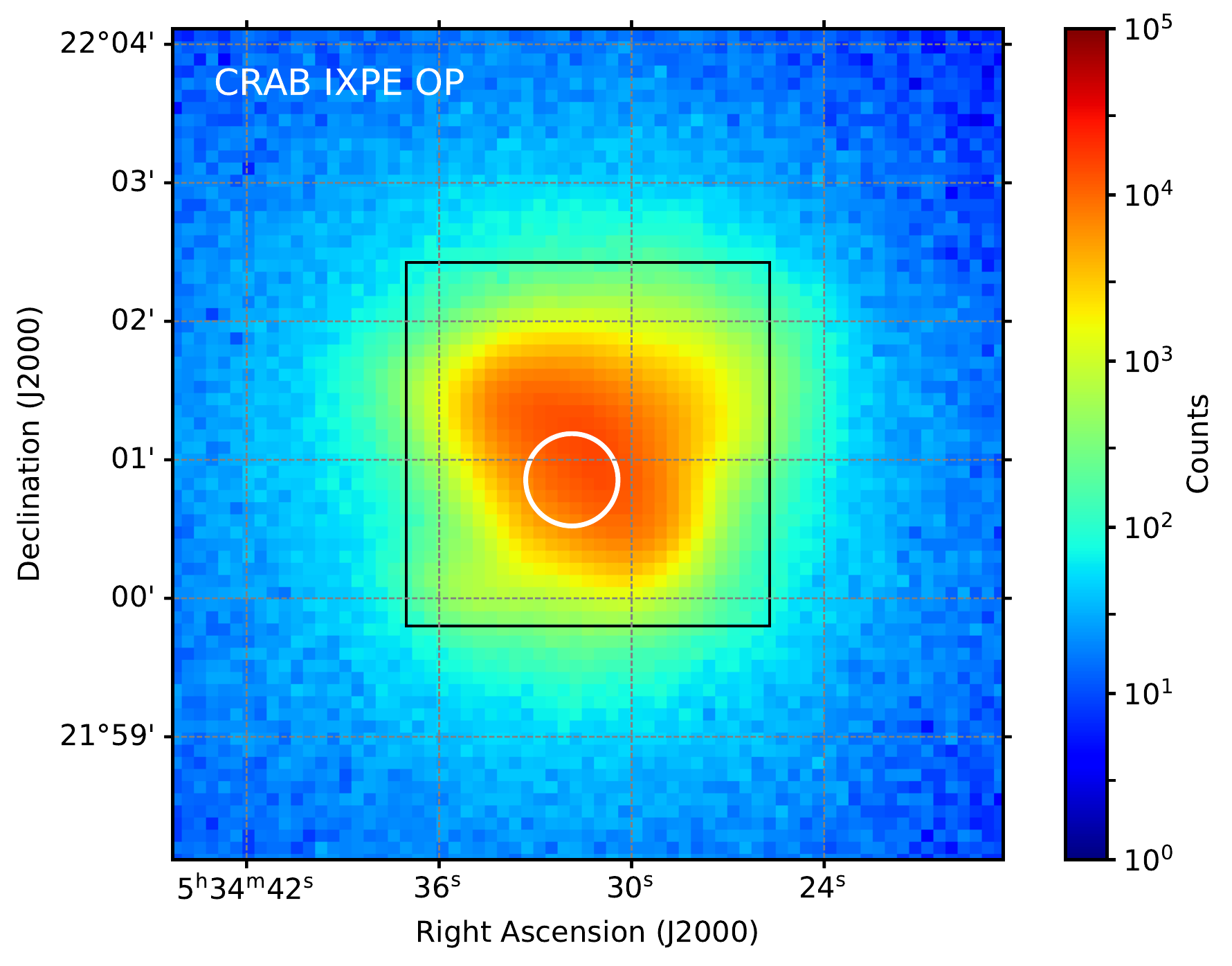}
\caption{\textbf{omparison of Chandra and IXPE images.} Top Left panel: Chandra image (ObsId235309) (intensity map) of the Crab PWN in the [2-8] keV energy range. Top Right panel: cartoon of the jet-torus structure indicating the main features observed in X-rays. Bottom Left panel: Total IXPE count map in the [2-8] keV energy range. Bottom Right panel: OP (see the Supplementary Materials for the definition of its phase range) only IXPE count map in the [2-8] keV energy range. The white circular region of 20'' radius is the one used to do the phase-resolved polarimetry of the PSR. The black box represents the region corresponding to the Chandra image.}\label{fig:2}
\end{figure}

\begin{figure}[H]%
\centering
\includegraphics[width=0.9\textwidth]{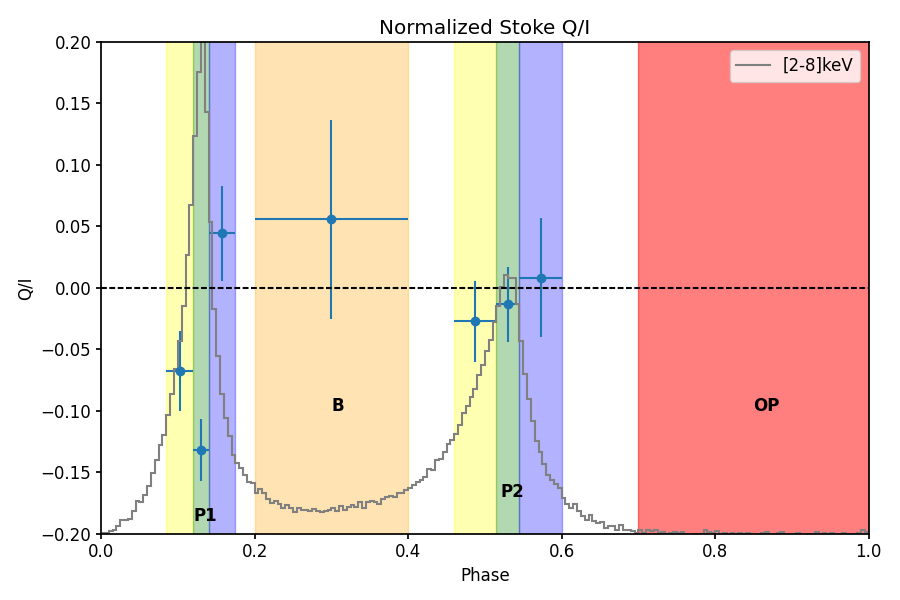}\\\includegraphics[width=0.9\textwidth]{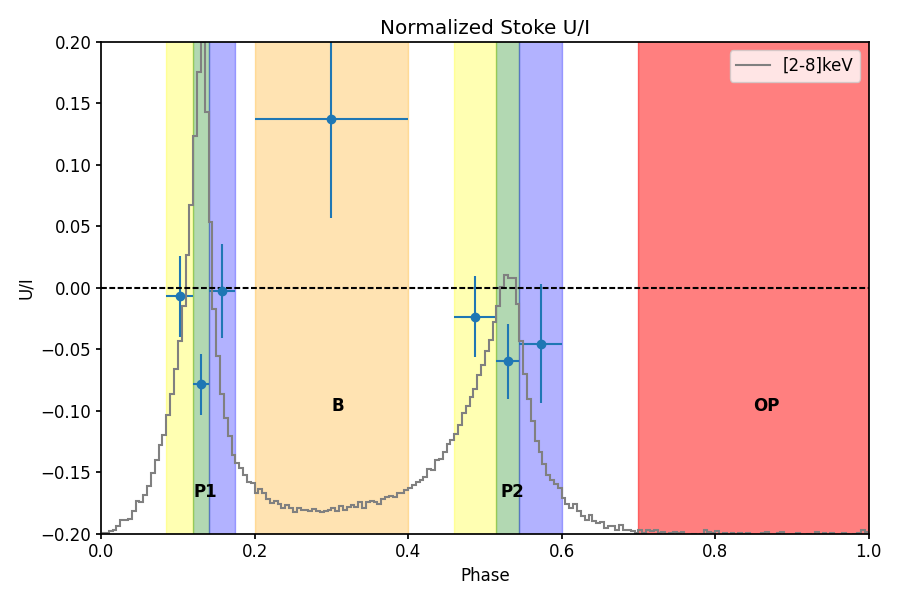}
\caption{\textbf{Polarization properties of the Crab PSR.} Normalized Stokes Parameters with their 1$\sigma$ errors for the OP subtracted emission of the Crab PSR in the [2-8]~keV energy bands (blue crosses), overlaid with the phase bins of interest: OP (red), P1 and P2 centers (green), left wings (yellow) and right wings(blue), and B (orange) (see the Supplementary Materials for the definition of the various phase ranges). Overlaid is the OP subtracted PSR light curve (counts) in the [2-8]~keV energy band, normalised and rescaled to the range of the y-axis (grey solid line).}\label{fig:3}
\end{figure}

\begin{figure}[H]%
\centering
\includegraphics[width=0.45\textwidth, trim = 60bp 0 80 0bp, clip]{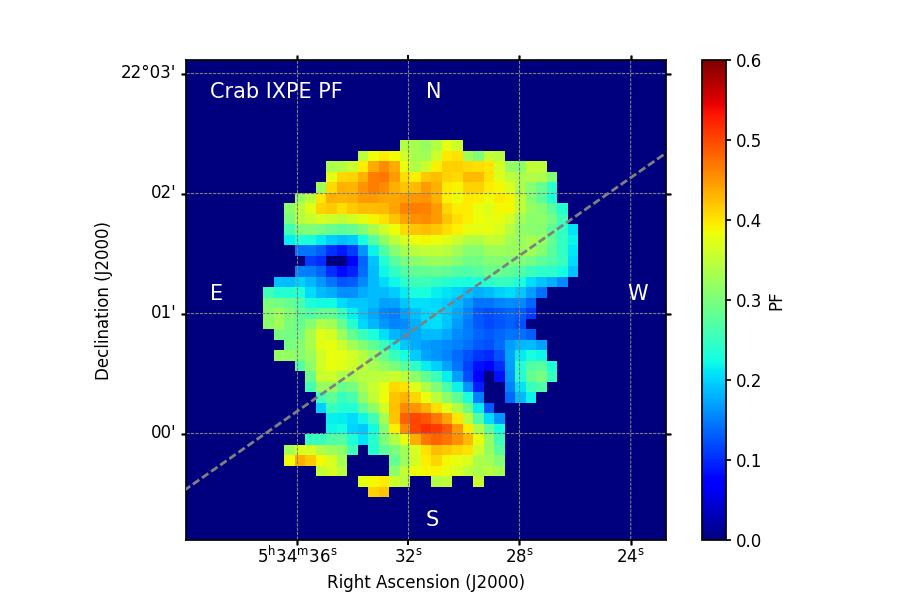}
\includegraphics[width=0.45\textwidth, trim = 80bp 0 60 0bp, clip]{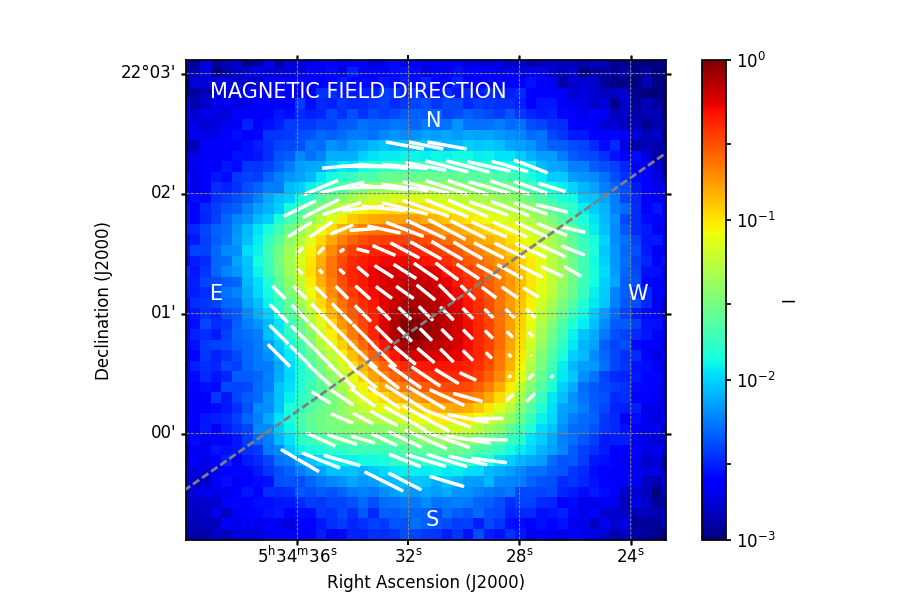}
\caption{\textbf{Polarized Structure of the Crab nebula.} Right panel: Total intensity map of the Crab PWN$+$PSR complex in the [2-8] keV energy band, overlaid with the reconstructed polarisation direction (magnetic field). Left panel: map of the polarized fraction (PD/100) cut above 5$\sigma$ confidence level. The gray dashed line id the nebular axis inferred from X-ray intensity maps \cite{Ng_Romani04a}. Overlaid are the Sky directions for ease of reference.}\label{fig:4}
\end{figure}

The Imaging X-ray Polarimetry Explorer (IXPE), the first mission devoted to spatially-resolved polarisation measurements in the X-rays \cite{Weisskopf22d,Weisskopf_Soffitta+22x}, was successfully launched on 2021 December 9. IXPE observed the Crab PWN \& PSR complex two times between February 21st and March 7th 2022 for a total on-source time of  $\sim$92ks. Data were extracted and analysed according to standard procedures: \texttt{HEASOFT 6.30.1}\footnote{https://heasarc.gsfc.nasa.gov/docs/software/heasoft/}  was used to perform barycenter correction, with the \texttt{BARYCORR FTOOL}, using the most recent optical coordinates from the Gaia Data release 3\footnote{https://www.cosmos.esa.int/web/gaia/dr3}, the DE421 JPL ephemeris and the ICRS reference frame (see Table S1 in the Supplementary Material). ixpeobssim V26.3.2 was used to do energy calibration, detector WCS correction, aspect-solution corrections, and all further (unweighted) analysis \cite{Pesce-Rollins_Lalla+19e,Baldini_Bucciantini+22c}, including phase folding at the derived ephemeris. We also performed a coeval observation of the Crab PWN with the CHANDRA Satellite (ObsId 23539, see the   Methods for further details).\\
\\
In Figure~\ref{fig:1} (see also Table 1) we show the polarised properties of the PWN derived by integrating all emission in a region within 2.5' from the PSR. Background contamination within this region is negligible. There is a significant change in the PA between the low [2-4]~keV and high [4-8]~keV energy band (true also for the OP emission - see Table S2 of the Supplementary Material for a definition of the OP phase range). The change in
PD is less significant, with the higher energy band being slightly more polarised. The OP phase emission is marginally more polarised than the total PWN \& PSR emission, as
already suggested by a similar analysis of the OSO-8 data \cite{Weisskopf_Silver+78a}, while there is no evidence for even a marginal change in PA, suggesting that the PSR has a net low level of
polarisation, acting mostly to reduce the total level of polarisation. The contribution of the PSR
unpulsed (DC) emission \cite{Tennant_Becker+01a} to the total OP emission, is estimated to be less than 1\%, and we
can safely assume that the OP is mostly of nebular origin. The OP polarisation angle is $\sim$145$^\circ$ ,
larger by $\sim$20$^\circ$  with respect to the values reported in the literature for the PWN symmetry
axis, derived from fitting the X-ray jet-torus intensity maps \cite{Weisskopf22d,Ng_Romani04a}. It is also smaller by
$\sim$10$^\circ$ than the previous OSO-8 measurements at a statistically significant level (see the
 Supplementary Material Figure S1). Such discrepancies might simply reflect the variability of
the PWN, where structures are known to change in shape and location over a typical timescale
of a few years \cite{Schweizer_Bucciantini+13a}.\\
\\
In Figure~\ref{fig:2} we show an intensity map of the Crab PWN from the coeval CHANDRA ObsId 23539,
the IXPE count map for the PWN$+$PSR complex in the [2-8]~keV energy range, and the IXPE count map in the same energy range, but computed just for the OP emission.\\
\\
For the phase-resolved analysis of the PSR we take events in the range [2-8]~keV and within
20'' from the PSR itself to limit the PWN contamination. The Stokes parameters of the OP
emission have been subtracted (see the   Methods for the exact definition of
the off pulse in terms of pulse phase, as well as the other phase bins). In Figure~\ref{fig:3} we plot the OP-subtracted light curve, in 200 equally spaced phase bins. For the polarisation analysis of the PSR emission we opted for a variable phase binning, focusing on the peaks and bridge, in order to get a finer sampling near P1 and P2. Figure~\ref{fig:3} shows PSR normalised Stokes parameters U/I and Q/I, for the phase bins of interest (see the  Supplementary Material Table S2 for further details). The OP emission in the pulsar aperture has Q/I = -0.0106 $\pm$ 0.008, U/I = -0.241 $\pm$ 0.008, corresponding to PD = 24.1\% $\pm$ 0.8\% and PA = 133.6$^\circ$  $\pm$ 1.0$^\circ$ . This is significantly more polarised than the OP emission for the entire PWN, and the PA is  $\sim$10$^\circ$  smaller, indicative of a spatial variation of the polarisation properties. The only phase bin showing a polarisation
above the 3$\sigma$ confidence level is the center of P1 in the phase range [0.12,0.14], where the OP
subtracted emission has Q/I = -0.132 $\pm$ 0.025, U/I = -0.079 $\pm$ 0.025, corresponding to
PD = 15.4\% $\pm$ 2.5\% and PA = 105$^\circ$  $\pm$ 18$^\circ$ . There is no significant change of the polarisation
properties of this phase bin with energy. Rapid PA variation might suppress the polarisation in these bins. The total PSR normalised Stokes parameters are Q/I
= -0.018 $\pm$ 0.019 and U/I = -0.019 $\pm$ 0.019, confirming that the integrated PSR contribution serves only to to reduce the polarisation of the entire complex.\\
\\
In Figure~\ref{fig:4} we show the total PWN \& PSR map of PD in the [2-8]~keV energy
range, obtained by smoothing the maps of Stokes parameters with a Gaussian kernel, and cut
at the 5$\sigma$ significance level (see the Supplementary Material Figure S2), together with
an intensity map where we have overlaid the polarised magnetic field direction. The overall
polarisation pattern agrees with the general expectation for PWNe, where the synchrotron emission takes place in the (mostly) toroidal
magnetic field, originating from the pulsar wind and compressed in the nebula, which sets the symmetry axis of the jet-torus structure.  The observed polarisation and emission patterns arise from the interplay of the magnetic field geometry and bulk motion of the relativistic plasma within the nebula itself, depending also on the inclination of the nebular axis with respect to the line of sight.  It is indeed the presence of bulk motions directed toward and away from the observer that creates the various bright arc-like features and makes the front side of the torus brighter than the back. 
The results shown in Figure~\ref{fig:4} agree with this picture, assuming a symmetry axis inclined in the plane of the sky as derived from X-rays \cite{Ng_Romani04a}. The direction of the inferred magnetic field broadly follows the shape of the emission torus (which extends also on the back but is fainter due to Doppler de-boosting). There are two unpolarised regions at the NE and SW edges of the main torus, where the polarised direction varies rapidly within the point spread function. The overall PD map shows a far stronger level of asymmetry with respect to the PWN axis than the total intensity map, indicating possibly large variations in the amount of magnetic turbulence within the PWN, or major distortions of the magnetic field structure in the fainter outer regions. The more polarised regions are not found in the center of
the PWN, where there is a marginal contribution from the PSR that lowers the PD,
but north and south of the main torus, in regions that do not correspond to any bright feature in
X-ray. Based on smoothed maps the peak PD in the Northern region is found to be
46\%, with a PA of 163$^\circ$ , at 25$\sigma$ significance, while the peak in the southern region
has PD of 51\%, with a PA 156$^\circ$ at 20$\sigma$ significance (see Methods Figure S3). Considering instead two circular regions of 15'' radius centered on the polarisation maxima, we found for the northern region integrated quantities: Q/I = 0.37 $\pm$ 0.01 and U/I = -0.25 $\pm$ 0.01, corresponding to PD 45\% $\pm$ 1\% and PA 163.3$^\circ$  $\pm$ 0.8$^\circ$ , at 35$\sigma$ significance with an MDP at 99 confidence level of 0.04, while the Southern region has Q/I = 0.30 $\pm$ 0.02 , U/I = -0.37 $\pm$ 0.02, corresponding to PD 47\% $\pm$ 2\% and PA = 154$^\circ$ $\pm$ 1$^\circ$ , at 27$\sigma$ significance with an Minimum Detectable Polarisation (MDP) at 99\% confidence level of 0.05. These regions are far enough from the PSR that its depolarising contribution is negligible. We caution the reader that due to error in the reconstruction of the photons absorption point in the Gas Pixel Detector (GPD), polarisation leakage can contaminate the Stokes maps, leading to spurious polarisation patterns (this has no effect on integrated or OP subtracted values). Preliminary estimates based on Montecarlo simulation of the GDP response, indicate that this effect can at most be as high as 10\% in the outer regions of the PWN, and does not alter significantly our overall findings.

\section{Discussion}\label{sec3}

We report here the first phase- and space-resolved soft X-ray polarised observation of the Crab PSR and PWN. Our results show that the total polarisation of the pulsed signal is negligible. The consistency with optical polarisation measure is marginal. Only the core of the main peak was found to be significantly polarised. Marginal evidence for polarisation, below 3$\sigma$, in other phase bins are also reported. The low average polarization is in contrast with the vast majority of the existing PSR models \cite{Dyks_Harding04a,Petri_Kirk05a} 
which typically predict polarisation fraction in the pulsed emission of 40-80\%. The model polarisation is generally especially high in the B region. The peaks which are believed to be caustics are	typically de-polarised via rapid PA sweeps. In contrast we find our	highest PD in the core of P1. Moreover a simple PA swing does not seem capable by itself to explain the presence of a highly polarized core in P1 surrounded by low polarization wings, unless the PA swings much faster than in optical. Intrinsic depolarization is most likely required.  Analytical striped-wind emission models suggest possible lower polarization in B, but also predict a fully unpolarized P1 \cite{Petri13a}. However recent models, focused on emission in the wind and outer magnetosphere, based on numerical magnetospheric solution have shown that the polarization signatures are highly sensitive to the location and geometry of the emission region \cite{Ceritti_Mortier+16a,Harding_Kalapotharakos17a}. Low integrated polarization suggests that the emission region should be close to or beyond the Light Cylinder \cite{Harding_Kalapotharakos17a}. However, none of the current models include important physical ingredients: micro-turbulence, which is likely present in the wind current sheet \cite{Cerutti_Philippov+20a}, and could lead to significative depolarization; short time-scale variability which manifests as timing noise \cite{Scott_Finger+03a} and could lead to potential depolarization for long time integration.\\
\\
A detailed comparison with previous measures, typically in higher energy ranges, or with optical data would benefit from a better statistics, and would require  further modeling/extrapolation (to account for changes in the pulse shape with energy) and goes beyond the scope of this work. Previous measures in hard X-rays for the PD of the integrated pulsed emission have values typically higher  $\sim$20\%-30\%.  These are well above our IXPE MDP for the pulsed emission. Our measure is in clear contrast with these high energy PD estimates. Note however, that the more recent PolarLight narrow band [3-4.5]~keV measures \cite{Long_Feng+21a} are consistent with our findings that the PSR emission is likely unpolarized. A strong decrease in the PD, from the	optical to the soft X-ray (with a possible recovery to large PD in the hard X-ray) is not expected in existing modeling, which primarily 	relies on geometry of the emission zone to determine the polarization	(e.g. \cite{Harding_Kalapotharakos17a}); as noted above, additional physical effects will be required to accommodate the IXPE data. \\
\\
The PWN shows a polarisation pattern that is consistent with a predominantly
toroidal magnetic field, extending well beyond the observed location of the X-ray torus. For synchrotron radiation this is consistent with general expectation from MHD modelling of this
source \cite{Bucciantini_del-Zanna+05a,Nakamura_Shibata07a,Porth_Komissarov+14a,Olmi_Del-Zanna+16a,Bucciantini_Bandiera+17a}. We found however that the mostly symmetric PA  (i.e. magnetic field) m pattern is associated with large asymmetries in the PD, likely indicating variations in the level of
turbulence inside the PWN. Such level of asymmetry is similar to, but stronger than, that
seen in the intensity maps, and reflects a similar trend found in optical polarisation images \cite{Hester08a}.\\
\\
The magnetic axis of the PWN, derived by taking the symmetry axis of the magnetic field pattern, is found to be $\sim140^\circ$, larger by about 15◦ than estimates based
on fitting axisymmetric structure to the torus intensity \cite{Ng_Romani04a}. It is also possible to estimate the
inclination of the magnetic axis with respect to the plane of the sky: we found it to be $sim60^\circ$,
in agreement with previous estimates. While the average PD $\sim$20\% agrees with
previous measures, the PA differs in a statistically significant way from other estimates, reflecting the spatial variation of the  PD, or possible temporal variability. The spatially resolved PD reaches a maximum of $\sim$46-50\%. This is about two times larger than expected from simple predictions based on synchrotron turbulent modelling of the Crab Torus and Inner Ring luminosity profiles, calibrated on the OSO-8 results \cite{Bucciantini_Bandiera+17a}. More sophisticate 3D models (lacking however micro-turbulence) can give PD close to the theoretical maximum $\sim$0.7 \cite{Porth_Komissarov+14a}, with higher values typically in the South-West region, but in general the prediction is for polarised patterns quite symmetric with respect to the nebular axis, unlike what was found. This suggests that the level and development of turbulence within the nebula, is not as strong as predicted and much patchier in its spatial distribution. While the lower level of polarization close to the center of the PWN is easily explained by summed emission from a wide range of PA in the central resolution elements, the increase of the PD with distance at the rim of the torus suggests the presence of a highly ordered magnetic layer at the edge of the torus itself [the ratio of the energy in the turbulent versus ordered magnetic field components should be about a factor 2 smaller than in the core of the torus \cite{Bandiera_Petruk16a}]. This differs from what is seen in optical where higher polarization is found in inner features \cite{Moran_Shearer+13a}, suggesting that optical and X-ray emitting particles might be accelerated in different locations and sample different regions of the nebula as previously suggested \cite{Schweizer_Bucciantini+13a}. The fact that the PD (which depends on the ratio of magnetic energy in the turbulent to ordered components) is far more asymmetric, with respect to the nebular axis, than the intensity (which depends on the total turbulent plus ordered magnetic energy density) suggests that the level of turbulence anti-correlates with the strength of the ordered component of the magnetic field. This is what one would expect if turbulence was driven by the growth of instabilities, like Rayleigh-Taylor, which are suppressed by stronger fields \cite{Bucciantini_Amato+04a}. If this is correct we should expect that more highly polarized PWNe (less turbulent systems) should show a stronger toroidal patterns with with smaller degree of Rayleigh-Taylor induced patchy de-polarisation and intensity enhancement [see e.g. the recent IXPE observation of the Vela PWN (Fei et al. 2022, submitted to Nature)]. \\
\\
The IXPE polarisation results indicate that present modeling lacks	physical ingredients needed to explain the low pulsar polarisation seen at most phases. The substantial spatial variation of the PD in 	the nebula also indicates that effects are missing even in the most advanced 3D relativistic Magneto-Hydrodynamical models; MHD turbulence seems likely to be important in both cases.


\backmatter
\bmhead{Supplementary information}
Supplementary material includes Figures S1 to S3, and Tables S1 to S2.

\bmhead{Acknowledgments}
The Imaging X-ray Polarimetry Explorer (IXPE) is a joint US and Italian mission. The US contribution is supported by the National Aeronautics and Space Administration (NASA) and led and managed by its Marshall Space Flight Center (MSFC), with industry partner Ball Aerospace (contract NNM15AA18C). The Italian contribution is supported by the Italian Space Agency (Agenzia Spaziale Italiana, ASI) through contract ASI-OHBI-2017-12-I.0, agreements ASI-INAF-2017-12-H0 and ASI-INFN-2017.13-H0, and its Space Science Data Center (SSDC), and by the Istituto Nazionale di Astrofisica (INAF) and the Istituto Nazionale di Fisica Nucleare (INFN) in Italy. This research used data products provided by the IXPE Team (MSFC, SSDC, INAF, and INFN) and distributed with additional software tools by the High-Energy Astrophysics Science Archive Research Center (HEASARC), at NASA Goddard Space Flight Center (GSFC). The research at Boston University was supported in part by National Science Foundation grant AST-2108622. Part of the French contributions is supported by the Scientific Research National Center (CNRS) and the French spatial agency (CNES).

\bmhead{Author contributions}
NB led the data analysis and the writing of the paper. RF, MB, JR, LP, FM contributed to data analysis and data calibration. NDL, CS, NO, TK, TM, SG, EW contributed to data analysis and results interpretation. MCW, MN, SS, EDOW, FX, JH, RWR, PT, AP, HLM contributed to text revision, and data interpretation. LB and MPR contribute to software development. The remaining members of the IXPE collaboration contributed to the design of the mission, to the calibration of the instrument, to define its science case and to the planning of the observations. All authors provided inputs and comments on the manuscript.
\bmhead{Competing interests}
Authors declare that they have no competing interests.
\bmhead{Availability of data and materials}
Data of the Crab pulsar and nebula observation are available in the HEASARC IXPE Data Archive (https://heasarc.gsfc.nasa.gov/docs/ixpe/archive/). 

\newpage
\begin{table}[h]
\begin{center}
\caption{\textbf{Global polarisation properties of the PSR+PWN complex.} Normalized Stokes parameters, Polarised Degree and Angle for various energy and phase ranges (in brackets the 1$\sigma$ errors). The OP is in the [2-8]~keV range. The significance (Sig) is given as the ratio of PD over its 1$\sigma$ error. See Figure~\ref{fig:1}}\label{tab:1}%
\begin{tabular}{@{}llllll@{}}
\toprule
Selection & Q/I  & U/I & PD[\%] & PA[deg] & Sig\\
\midrule
PSR+PWN [2-8] keV  & 0.177(0.0019)  & 0.068(0.0019) & 19.0(0.19) & 145.5(0.29) & 99\\
PSR+PWN [2-4] keV  & 0.168(0.0019)  & 0.081(0.0019) & 18.7(0.19) & 147.8(0.29) & 100\\
PSR+PWN [4-8] keV  & 0.199(0.0038)  & 0.037(0.0039) & 20.2(0.38) & 140.2(0.55) & 53\\
PWN (OFF Pulse)    & 0.189(0.0036)  & 0.073(0.0036) & 20.2(0.36) & 145.6(0.51) & 57\\
\end{tabular}
\end{center}
\end{table}

\clearpage
\newpage


\bibliography{sn-bibliography}

\newpage
\section*{Methods}
\subsection{Observations and Data Analysis}

The Imaging X-ray Polarimetry Explorer (IXPE), is a NASA mission in partnership with the
Italian space agency (ASI), launched in December 9 2021. As described in detail elsewhere
[\cite{Weisskopf22d,Weisskopf_Soffitta+22x} and references therein], the IXPE Observatory includes three identical X-ray telescopes (DUs), each comprising a Wolter-I X-ray mirror assembly (NASA-provided) with with angular resolution (half-power diameter) of 19'' (DU1), 26'' (DU2), 28'' (DU3), and a polarisation-sensitive pixelated detector (GDP, ASI-provided), with a typical energy-dependent dead-time of
$\sim$1.1 ms. This allows one to measure the energy, arrival direction, arrival time and linear polarization of the detected X-ray signal, which are all reconstructed from the photo-electron track
shape using moment analysis. The IXPE energy range is the [2-8] keV band, with a total effective area of 590~cm$^2$ at 4.5~keV. The modulation factor (the amplitude of the modulation of the reconstructed photo-electrons angle distribution for a
100\% polarised source) ranges from $\sim$15\% at 2~keV up to $\sim$60\% at 8~keV \cite{Weisskopf_Soffitta+22x}.\\
\\
The Crab PWN and PSR were observed twice: the first time from 2022-02-21 UTC16:13:32 to 2022-02-22 UTC18:46:37, the second from 2022-03-07 UTC00:14:20 to 2022-03-08 UTC02:40:02, for a total 92363 seconds of ONTIME exposure (the total exposure as obtained from the sum of the good time intervals), and 85062 seconds of total LIVETIME (the total amount of time the CCD was actively observing a source; it excludes the time it takes to transfer charge from the image region to the frame store region).\\
\\
The polarisation analysis was performed on publicly available level 2 event list files. These
were corrected to account for a few calibration issues that have emerged during flight operations. The World Coordinate System (WCS) was corrected to account for the small offset
among the various units, registering the pointing solution in order to center the intensity peak of
each unit on the PSR position at RA = 5h:34m:31.86s, DEC = 22:00:51.3. The time dependent
charge-to-energy conversion was reconstructed for each units using the two onboard calibration
sources at 1.7 keV (Si K$\alpha$) and 5.9 keV ($^{55}$Fe$\rightarrow$ $^{55}$Mg with following K$\alpha$ emission) \cite{Ferrazzoli_Muleri+20a}.
Spurious offsets in the pointing solution (aspect solution),
associated to the switch between different star trackers during orbit, were identified looking at
the time variations of the count rate in a set of background sky regions, and later filtered out
by removing the affected time intervals. This results in a loss of counts smaller than 2\% and
an effective new on-source time of $\sim$91ks. At the time of the Crab observation, the optical
axis of the mirror system relative to the star trackers had not yet been accurately determined
and was not yet compensated (by offset pointing). Consequently, the Crab PSR was about
2.8' off-axis with respect to the mirror system. This precluded accurate computation of Image Response Functions (IRFs)-energy dependent vignetting and energy dependent exposure
maps—necessary for a correct spectral analysis (the mirror-system–star-tracker offset has now
been accurately determined and compensated, such that future observations will place the image close to the mirror-system optical axis. Furthermore, it may be possible, in the future, to recalibrate the IRFs), and for computing correct counts rates. However this should not affect polarisation measures which comes from flux ratios among Stokes parameters.       

\subsection{Timing Analysis}
We initially used the Jodrell Bank Crab PSR Monthly ephemeris (http://www.jb.man.ac.uk/pulsar/crab.html) \cite{Lyne_Pritchard+93a} to calculate the pulse phase
of each photon. However, the time span of the two IXPE observations requires two separate JB
solutions, and the alignment between the arrival time of the pulse in the two observations using
these ephemerides is visibly off ($\sim$0.02 in phase). Therefore, we determined a new ephemeris
by using the X-ray data alone, as follows. As a starting point, we used the JB montly ephemeris
of February in CGRO format (https://www.jb.man.ac.uk/$\sim$pulsar/crab/CGRO\_format.html) , but modifying the frequency and derivatives in order for them to refer to an epoch between the two observations:
\begin{eqnarray}
\nu_{\rm new} =& \nu_{\rm old}+\dot{\nu}_{\rm old}(T_{\rm  new}-T_{\rm  old}) +0.5 \Ddot{\nu}_{\rm old}(T_{\rm  new}-T_{\rm old})^2\\
\dot{\nu}_{\rm new} =& \Ddot{\nu}_{\rm old}(T_{\rm  new}-T_{\rm old})
\end{eqnarray}
where $\nu_{\rm old}$, $\dot{\nu}_{\rm old}$, and $\Ddot{\nu}_{\rm old}$ are the frequency, its first and second time derivative at $T_{\rm old}$, while $\nu_{\rm new}$, $\dot{\nu}_{\rm new}$d are the frequency and its first time derivative at $T_{\rm new}$. \\
\\
Then, we calculated times of arrival (TOAs) of the pulse using the HENphaseogram tool
distributed with \texttt{HENDRICS} (https://hendrics.stingray.science) \cite{Bachetti08a}. This tool folds the data in small fractions of the observation, creating a series of pulse profiles. Then, it calculates the misalignment between each of these profiles and a smoothed version of the folded profile from the whole observation, and transforms this misalignment into a TOA. The TOA refers to the maximum of the reference profile. As an output, the tool produces a parameter file and a TOA file in Tempo2 format \cite{Hobbs_Edwards+06a}. We loaded these approximate parameters and TOAs in the \texttt{pintk} graphical interface to \texttt{PINT} (https://nanograv-pint.readthedocs.io) \cite{Luo_Ransom+21a}, and fitted a new spin down solution that aligned the TOAs of the two observations. Then, we went back and calculated new TOAs, this time using the improved timing solution and, consequently, the better resolved total pulse profile that the solution provided, and fitted these TOAs again with PINT. We stopped this iterative procedure when the improvement in the fitting through PINT was smaller than the uncertainties. Finally, we calculated the closest TOA to the epoch chosen for the timing solution, and referred the frequency and derivatives to this time, using the above equation, in order to have a single number for the reference TOA and the PEPOCH of the timing solution for convenience. The new ephemeris is reported in Table S1 of the Supplementary Material. Note that, having chosen the reference time in between the February and March observations, the determination of the second time derivative of the period is highly uncertain, but  on the other hand the correction due to the second time derivative is not significative, and can potentially be neglected. Note moreover that the absolute time alignment of the pulse profile is not necessary for the analysis in this Paper. However we did verify that the X-ray pulse leads the radio pulse (as provided by the JB ephemeris) by $\sim$300 $\mu$s, consistent with past observations from other missions \cite[e.g.]{Kuiper_Hermsen+03a}.  The Crab pulsar is a young one and his timing can be quite noisy, to the paint that ephemeris cannot be extended beyond their range of validity.

\subsection{Polarization Analysis}

The polarisation analysis of both the PWN and the PSR was performed using the IXPE collaboration software \texttt{ixpeobssim V26.3.2}. ixpeobssim has been designed to act both as a simulation software and for data reduction. We opted for a more robust and established unweighted analysis, limited to the [2-8] keV energy range. Phase folding was performed with \texttt{xpphase}, while event selection was done using the \texttt{xpselect} tool. Polarization was computed with the \texttt{xpbin} tool and  \texttt{PCUBE} and \texttt{PMAPCUBE} methods.\\
\\
Being Stokes parameters additive,
their values relative to each event, calibrated for the known spurious modulation of the instrument, and corrected for the modulation factor \cite{Weisskopf_Soffitta+22x} , are summed over the spatial region and/or phase range of interest. ixpeobssim can compute all polarisation relevant quantities including their error and the significance. Deformation of the phase-resolved light curve (pulse shape) due to dead time, with respect to the dead-time corrected one, is estimated to be less than 3\% [difference between the dead time correction when the count rate is at maximum (P1) and the one when to the count rate at minimum (OP)], and was thus ignored. The net PSR's Stokes parameters in a given phase bin are obtained as follows: the I, Q and U values for the OP are subtracted from those of the pulse phase bin of interest (after scaling then to the phase bin width) to give a net, OP-subtracted, I,  Q and U.   Figure S2 of the Supplementary Material shows the maps of normalized Stokes parameters, of the total PWN plus PSR emission. The contribution of the lowly polarised PSR is hardly visible. We verified that the pattern, apart from counting noise, is the same if one considers just the emission in the OP phase range. While the U/I map shows a high level of symmetry with respect to the direction of the nebular axis inferred from the intensity maps \cite{Ng_Romani04a}, the one for Q/I is far more asymmetric (its symmetry axis is more aligned with the North-South direction). The inclination of the magnetic axis of the nebula was derived by fitting ellipses to the internal magnetic field structure as displayed in Figure 4.

\subsection{CHANDRA Observation}
The Crab PWN was observed by CHANDRA (OsbId23539) starting on 2022-03-15 at 11:32:23
UTC and ending on 2022-03-15 at 15:14:17 UTC, for a total of $\sim$ 10$^4$ seconds; due to telemetry saturation from the bright source, the effective exposure time was 1331 seconds. Data were processed with the \texttt{CIAO} package 4.14 using \texttt{CALDB 4.9.7}, with the \texttt{chandra repro mode=h} tool using default settings, and the [2-8] keV image was done with the \texttt{fluximage} tool, and later smoothed with a gaussian kernel using  \texttt{aconvolve kernelspec='lib:gaus(2,5,1,1,1)'}

\clearpage

\section*{Supplementary Material}

\begin{figure}[H]%
\centering
\includegraphics[width=0.85\textwidth]{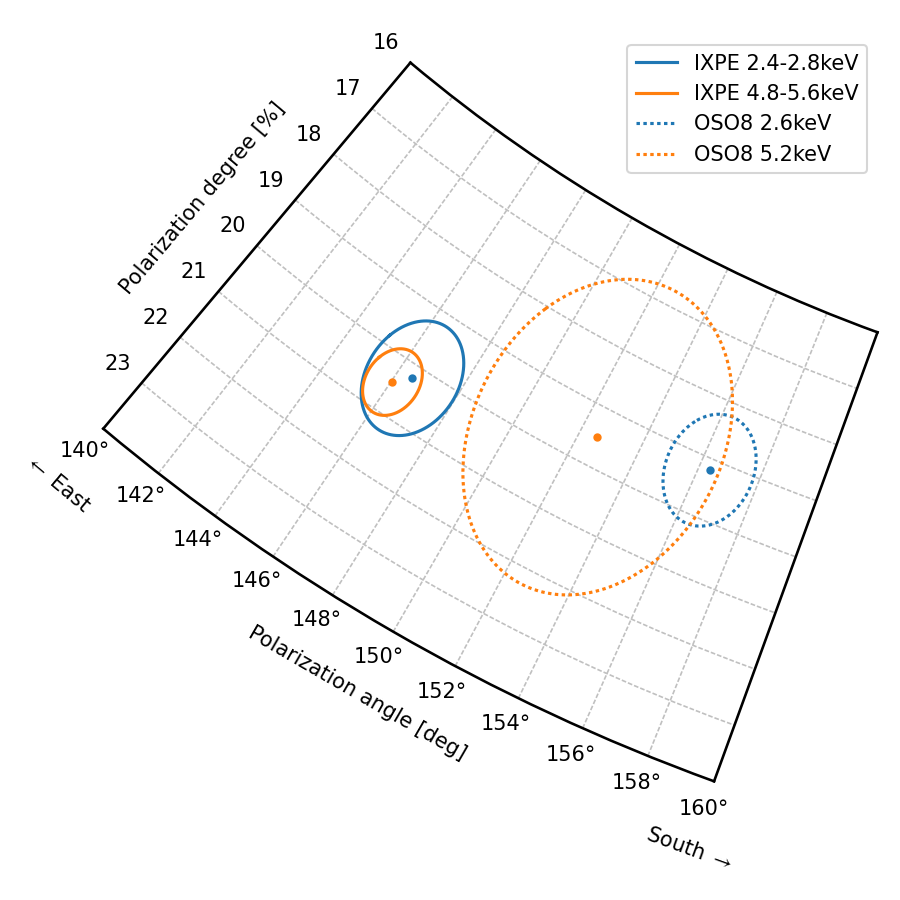}
\caption*{\textbf{Figure S1.} Comparison of the polarisation measured between IXPE and OSO-8 nearly 50 years before, both PSR subtracted, in a region of 110arcsec centered on the PSR, and covering the entire PWN. Contours are at 1$\sigma$ confidence (67\%). Spectral response of OSO-8 was strongly peaked in a narrow energy band, and for this reason we selected in the IXPE analysis only the events in two narrow energy bands bracketing the ones of OSO-8. The comparison shows that the polarisation degree measured by OSO-8 are fully in agreement with those measured by IXPE, whereas a rotation of the polarisation angle of about 10$^\circ$ is detected at more than 2$\sigma$ confidence for the lower energy band, and at about 1$\sigma$ level for the higher energy band.}
\end{figure}

\begin{figure}[H]%
\centering
\includegraphics[width=0.95\textwidth]{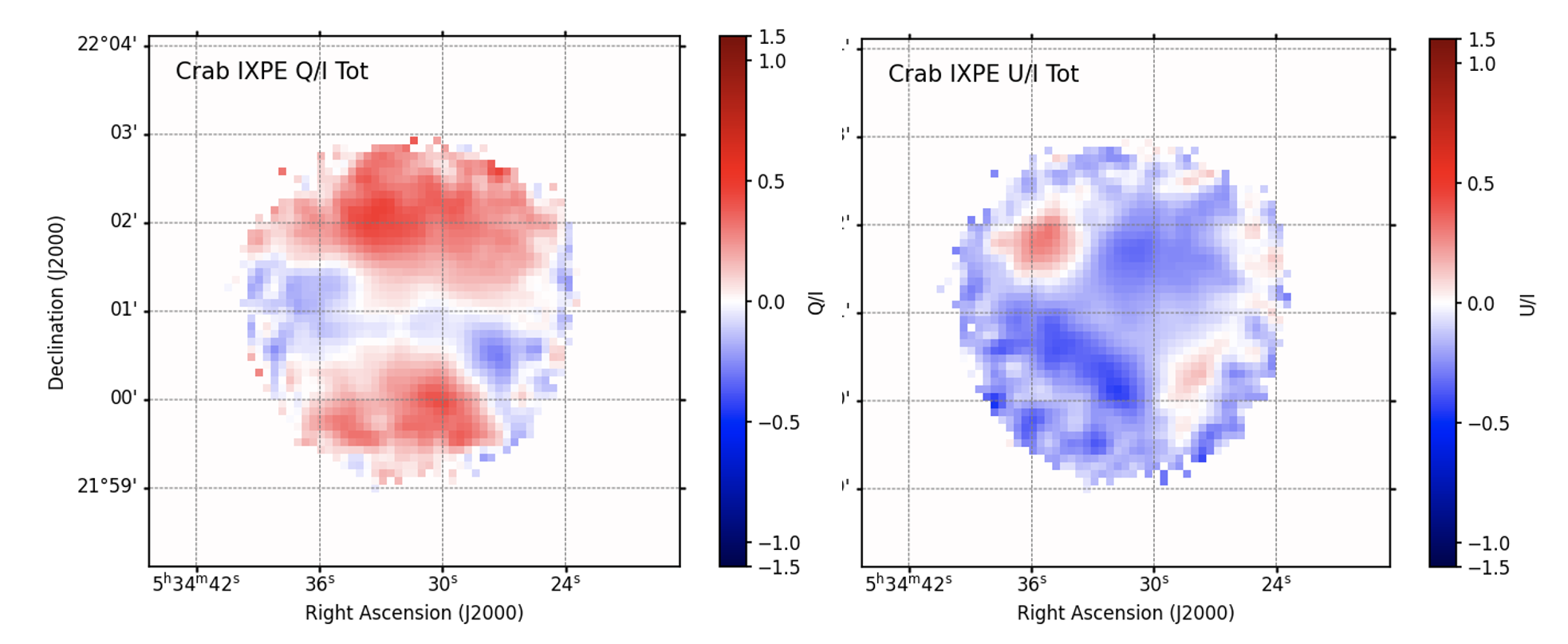}
\caption*{\textbf{Figure S2.} Maps of the total PWN \& PSR normalised Stokes Parameters Q/I and U/I, in the [2-8] keV energy band, smoothed with a Gaussian kernel, and limited to pixels having intensity above 0.3\% of the maximum. }
\end{figure}

\begin{figure}[H]%
\centering
\includegraphics[width=0.95\textwidth]{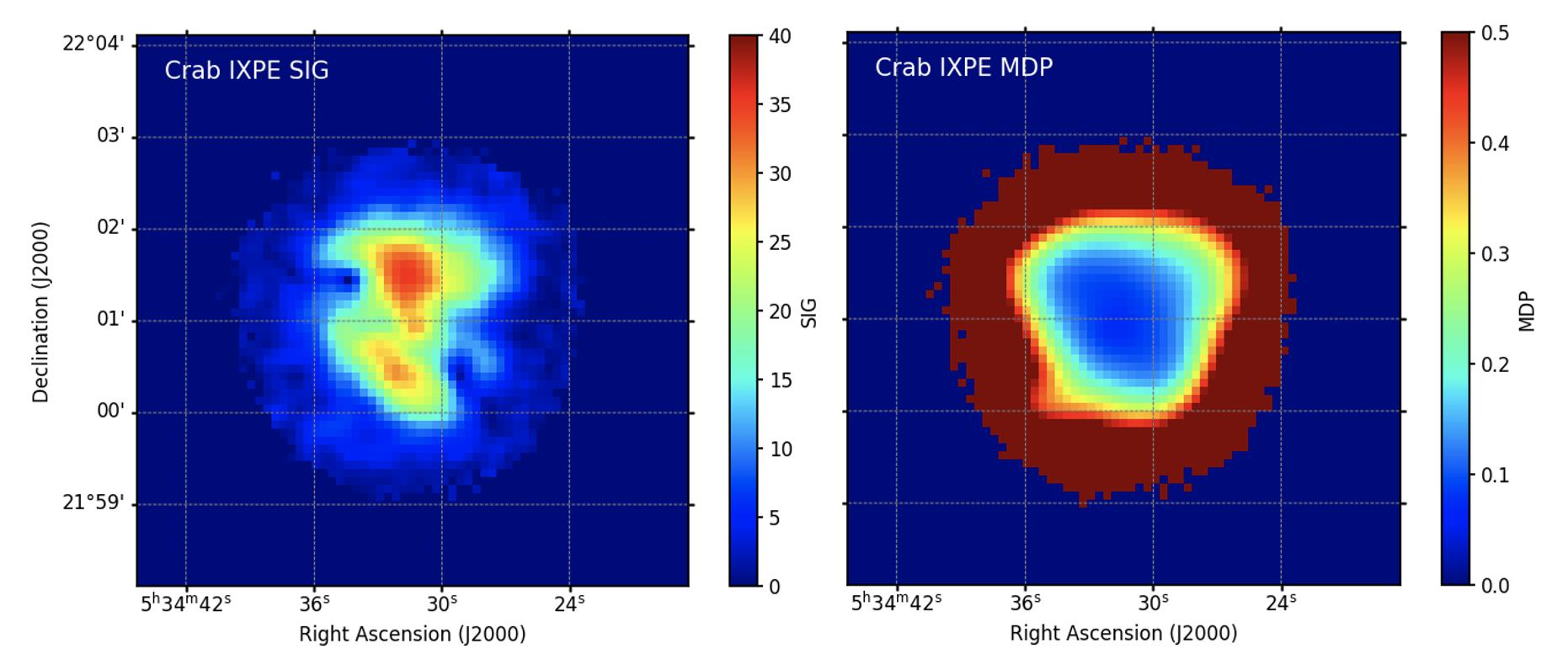}
\caption*{\textbf{Figure S3.} Left panel: map of the significance (ratio of polarised degree over its 1$\sigma$ uncertainty) in the [2-8] keV energy band, smoothed with a Gaussian kernel, and limited to pixels having intensity above 0.3\% of the maximum. Right panel: map of the Minimum Detectable Polarisation (MDP) at 99\% confidence level in the same energy band.}
\end{figure}

\begin{table}[h]
\begin{center}
\caption*{\textbf{Table S1} PSR ephemeris as determined from IXPE events time of arrival. F0 is the period, F1 the period time derivative, F2 the period second time derivative, all at PEPOCH. In brackets the 1$\sigma$ errors on the last digit. }
\begin{tabular}{@{}ll@{}}
\toprule
PEPOCH (MJD) & 59639.358974437835272\\
F0 (s) & 29.5866194471(4)\\
F1 & -3.67445(8)e-10 \\
F2 (s$^{-1}$) & -4(4) e-20 \\
\end{tabular}
\end{center}
\end{table}

\begin{table}[h]
\begin{center}
\caption*{\textbf{Table S2.} OP subtracted (and OP) polarisation properties of various phase bins (a phase offset has been added to shift P1 at phase 0.13, for ease of visualization). In brackets the 1$\sigma$ errors. See also Figure 3.}%
\begin{tabular}{@{}llll@{}}
\toprule
Phase Bin & Phase Range & Q/I  & U/I \\
\midrule
P1 left wing &	[0.085-0.120] &	-0.068(0.033) &	-0.007(0.033)\\
P1 center &	[0.120-0.140] &	-0.132(0.025) &	-0.079(0.025)\\
P1 right wing &	[0.140-0.175] &	0.044(0.038) &	-0.003(0.038)\\
P2 left wing &	[0.460-0.515] &	-0.027(0.033) & 	-0.023(0.033)\\
P2 center &	[0.515-0.545] &	-0.013(0.030) &	 -0.06(0.030)\\
P2 right wing &	[0.545-0.600] &	0.008(0.047) &	 -0.045(0.047)\\
Bridge &	[0.200-0.400] &	0.056(0.081) &	 0.137(0.081)\\
Off-Pulse &	[0.700-1.000] &	-0.01606(0.008) &	-0.241(0.008)\\
\end{tabular}
\end{center}
\end{table}

\end{document}